\documentstyle[12pt]{article}
\def\beq{\begin{equation}}
\def\eeq#1{\label{#1}\end{equation}}
\def\beqa{\begin{eqnarray}}
\def\eeqa#1{\label{#1}\end{eqnarray}}
\def\no{\nonumber}

\def\Boron{\hbox{${}^8$B}}
\def\Beryllium{\hbox{${}^7$Be}}
\def\Chlorine{\hbox{${}^{37}$Cl}}
\def\Gallium{\hbox{${}^{71}$Ga}}
\def\RSSM{\hbox{$R_{\rm SSM}$}}
\def\E{\hbox{$E_{\nu}$}}
\def\phisigmaKam{\hbox{$\phi\sigma(\nu_e e;\E)$}}
\def\phisigmaCl{\hbox{$\phi\sigma(\Chlorine;\E)$}}

\begin{document} 

\rightline{UTAPHY-HEP-9}
\rightline{4/18/94}
\vskip 4ex

\begin{center}
  {\large\bf Semi-empirical bound on the \Chlorine\ solar neutrino experiment}
  \\ \vskip 0.5in
  Waikwok Kwong and S. P. Rosen\\
  \vskip 1ex
  \it Department of Physics, University of Texas at Arlington\\
      Arlington, Texas 76019-0059
\end{center}

\vskip 2ex
\centerline{ABSTRACT}
\begin{quotation}
\noindent
The Kamiokande measurement of energetic \Boron\ neutrinos from the sun is used
to set a lower bound on the contribution of the same neutrinos to the signal
in the \Chlorine\ experiment. Implications for \Beryllium\ neutrinos are
discussed.
\end{quotation}

\vskip 1ex

Energetic \Boron\ neutrinos from the sun have been detected in the Kamiokande
experiment [1] at about one half the rate predicted by the Standard Solar
Model (SSM) [2]. These same neutrinos must also interact with the \Chlorine\
detector [3] and so it is important to understand their contribution to the
measured \Chlorine\ signal. By comparing this contribution to the total
signal, we can extract information about other parts of the solar neutrino
spectrum, especially \Beryllium.

We find that, even allowing for neutrino flavor oscillations, the Kamiokande
experiment imposes a bound on the \Chlorine\ signal that does not leave much
room for a significant contribution from \Beryllium\ neutrinos. This finding
is not inconsistent with the latest results from the \Gallium\ experiments
[4,5], and so we may refine the statement of the solar neutrino problem to
read: Where have all the \Beryllium\ neutrinos gone?

Since the basic physical process in the Kamiokande and \Chlorine\ experiments
are different, the former being neutrino--electron scattering and the latter
neutrino capture on \Chlorine, we must follow a semi-empirical method to
relate them to one another. In Kamiokande, the calculated signal involves a
convolution over $\phi(\E)$, the SSM spectrum of \Boron\ neutrinos with
energy \E, the differential cross section for scattered electrons
with kinetic energy $T$, and the electron resolution function $\theta(T,T')$
which represents the probability that $T$ will appears as $T'$ in an actual
measurement. We call this function \phisigmaKam\ and plot in Fig.~1 its
normalized shapes as a function of \E\ for two choices of $\theta(T,T')$: The
first is a Gaussian shape that closely approximates the actual experimental
resolution [6], the second is a $\delta$-function representing perfect
resolution, and both assume $7.5 \le T' \le 15$ MeV. Notice that because of
the experimental resolution, the first case has developed a significant tail
below the 7.5 MeV threshold. Only the first case with the experimental
resolution will be used for calculations below.

In the \Chlorine\ experiment, the relevant quantity is the product of
$\phi(\E)$ with the total capture cross section [7] for neutrinos of energy
\E\ on \Chlorine. We call this function \phisigmaCl\ and plot its normalized
shape also in Fig.~1. The integral of \phisigmaCl\ gives the \Boron\
contribution to the SSM signal in \Chlorine, \RSSM(\Beryllium;\Chlorine).

Comparing the normalized functions for the two experiments, we see that they
are remarkably similar to one another, especially at the high energy end. We
therefore write
\beq
  \frac{\phisigmaCl}{\int\phisigmaCl d\E}
  ~=~ \alpha\,\frac{\phisigmaKam}{\int\phisigmaKam d\E}~+~r(\E)~,
\eeq{eq1}
where $\alpha$ is a constant whose value is maximized subject to the condition
that the remainder function $r(\E)$ be everywhere positive. It turns out
that the largest value of $\alpha$ is 0.93, and so we obtain an inequality
\beq
  \phisigmaCl ~\ge~
  0.93\,\frac{\RSSM(\Boron;\Chlorine)}{\RSSM(\rm Kam)}\,\phisigmaKam~.
\eeq{eq2}

The next step of the argument is to note that the actual quantity measured
in these experiments involves the product of $\phi\sigma$ with an
electron-neutrino ``survival probability'' $P(\E)$ which, in general,
may be a function of the neutrino energy \E. If $P(\E)$
represents some, possibly energy-dependent, reduction of the \Boron\
spectrum, or an oscillation into a sterile neutrino, then we find from
Eq.~(\ref{eq2}) that
\beqa
  \int\phisigmaCl P(\E)\,d\E & \ge &
  0.93\,\frac{\int\phisigmaKam P(\E)\,d\E}{\RSSM(\rm Kam)}
  \,\RSSM(\Boron;\Chlorine) \no\\
\noalign{\noindent \rm or}
  R(\Boron;\Chlorine) & \ge & 0.93\,(0.50 \pm 0.08)\,(6.1 \rm~SNU) \no\\
  & = & (2.84 \pm 0.45) \rm~SNU~,
\eeqa{eq3}
where we have used the most recent result from the Kamiokande experiment [1].
This falls within the errors of the twenty-year average of the Davis value
[3]
\beq
  \langle R_{\rm Davis} \rangle = 2.32 \pm 0.23 \rm~SNU~,
\eeq{eq4}
but is somewhat on the high side. Note that the bound in Eq.~(\ref{eq3})
also holds in the simple case of a reduction of the total \Boron\ flux with
no change in the spectral shape.

Next, consider the case of oscillations of solar electron-neutrinos into
$\nu_{\mu}$ or $\nu_{\tau}$, or some combination thereof. The signal
observed in Kamiokande is then given by
\beq
  R(\hbox{Kam}) = \int \Bigl( \phisigmaKam P(\E)
  + [1-P(\E)]\,\phi\sigma(\nu_{\mu}e;\E) \Bigr) d\E~,
\eeq{eq5}
where we must now distinguish between the cross sections for
electron-neutrinos and muon- or tau-neutrinos. As is well known [7] the
latter cross section lies somewhere between 1/6 and 1/7 of the the former in
magnitude and is very similar in shape for energetic neutrinos. For our case
it is an extremely good approximation to set
\beq
  \sigma(\nu_{\mu}e;\E) = 0.148\,\sigma(\nu_e e;\E)~.
\eeq{eq6}
We can then rewrite Eq.~(\ref{eq5}) in the
form
\beqa
  \int \phi\Bigl(\sigma(\nu_e e;\E) - \sigma(\nu_{\mu}e;\E)\Bigr) P(\E)\,d\E
  &=& R(\hbox{Kam}) - \int \phi\sigma(\nu_{\mu}e;\E)\,d\E~, \no\\
\noalign{\noindent \rm or}
  0.852\,\int \phisigmaKam P(\E)\,d\E
  &=& R(\hbox{Kam}) - 0.148\,\RSSM(\hbox{Kam})~.
\eeqa{eq7}
{}From Eqs.~(\ref{eq2}) and (\ref{eq7}) and the Kamiokande data [1], we see
that the contribution of the \Boron\ neutrinos must be bounded in the case of
flavor oscillations by
\beqa
  R(\Boron;\Chlorine) & = & \int\phisigmaCl P(\E)\,d\E \no\\
  &\ge & 0.93\,\frac{\int\phisigmaKam P(\E)\,d\E}{\RSSM(\rm Kam)}\,
         \RSSM(\Boron;\Chlorine) \no\\
  & = & 0.93\,\frac{(0.50 \pm 0.08) - 0.148}{0.852}\,(6.1 \rm~SNU) \no\\
  & = & (2.34 \pm 0.53) \rm~SNU~.
\eeqa{eq8}

To show that the above argument really does provide lower bounds on the
\Beryllium\ neutrino contribution to the \Chlorine\ experiment, we consider
the special case in which, inspired by the non-adiabatic MSW solution [8],
we take the electron-neutrino survival probability to be [9]
\beq
  P(\E) = e^{-C/E_{\nu}}~,
\eeq{eq9}
where $C$ is a constant to be determined by fitting the Kamiokande data.
When there is either no oscillation, or oscillation into a sterile neutrino,
we find
\beq
  C = 6.9^{+1.8}_{-1.5} \rm~MeV \qquad\hbox{and}\qquad
  R(\Boron,\Chlorine) = 3.0 \pm 0.5 \rm~SNU~.
\eeq{eq10}
Allowing for neutrino oscillations, we find instead
\beq
  C = 8.8^{+2.6}_{-2.0} \rm~MeV \qquad\hbox{and}\qquad
  R(\Boron,\Chlorine) = 2.5 \pm 0.5 \rm~SNU~.
\eeq{eq11}
Both rates are larger than the corresponding lower bounds in Eqs.~(\ref{eq3})
and (\ref{eq8}) respectively.

When compared with the Davis result of Eq.~(\ref{eq4}), our bounds on the
energetic \Boron\ neutrino contribution in Eq.~(\ref{eq3}) and (\ref{eq8})
do not leave much room for the 1.8 SNU coming from all other sources, or
the 1.1 SNU from \Beryllium\ neutrinos alone. Indeed, the contribution from
all other sources, call them $X$, is given in the two cases we have
considered by
\beq
  R(X,\Chlorine)~\le~\left\{
  \begin{array}{l}
    -0.52 \pm 0.51 \rm~SNU~~~(no~oscillations), \\
    -0.02 \pm 0.58 \rm~SNU~~~(with~oscillations).
  \end{array}\right.
\eeq{eq12}
At the 95\% confidence limit, this means
\beq
  R(X,\Chlorine)~\le~\left\{
  \begin{array}{l}
    0.32 \rm~SNU~~~(no~oscillations), \\
    0.93 \rm~SNU~~~(with~oscillations).
  \end{array}\right.
\eeq{eq13}
Assuming that the \Beryllium\ contribution is approximately 1.1/1.8, or 60\%
of this, we find it to be:
\beq
  R(\Beryllium,\Chlorine)~<~\left\{
  \begin{array}{l}
    0.20 \rm~SNU~~~(no~oscillations), \\
    0.57 \rm~SNU~~~(with~oscillations).
  \end{array}\right.
\eeq{eq14}

To pursue this line of argument further, we can set lower bounds on the
contribution of the \Boron\ neutrinos to the \Gallium\ experiments.
Replacing the absorption cross section of \Chlorine\ by that of \Gallium\
everywhere [10], we obtain an inequality similar to Eq.~(\ref{eq2}) but with
$\alpha = 0.81$. The bounds on the \Boron\ contribution to the \Gallium\
experiments are
\beq
  R(\Boron,\Gallium)~\ge~\left\{
  \begin{array}{l}
    5.7 \pm 0.9 \rm~SNU,~~~(no~oscillations), \\
    4.7 \pm 1.1 \rm~SNU,~~~(with~oscillations).
  \end{array}\quad\right.
\eeq{eq15}
The corresponding values in the $e^{-C/E}$ model,
\beq
  R(\Boron,\Gallium) ~=~ \left\{
  \begin{array}{l}
    6.6 \pm 1.1 \rm~SNU,~~~(no~oscillations), \\
    5.5 \pm 1.3 \rm~SNU,~~~(with~oscillations),
  \end{array}\quad\right.
\eeq{eq16}
are again larger than their counterparts in Eq.~(\ref{eq15}).

Combining the bounds of Eq.~(\ref{eq15}) with the latest \Gallium\ results
[4,5],
\beqa
  R(\Gallium) &=& \left\{\begin{array}{ll}
                           79 \pm 12 \rm~SNU, & \rm GALLEX \\
                           73 \pm 19 \rm~SNU, & \rm SAGE
                         \end{array}\right. \no\\
              &=& 77 \pm 10 \rm~SNU,~~~(combined)
\eeqa{eq17}
we find an interesting situation, namely that the sum of the signals from pp
neutrinos, \Beryllium\ neutrinos, and other non-\Boron\ sources is very
close to the SSM prediction of 71 SNU for pp neutrinos alone:
\beq
  R(\Gallium) - R(\Boron,\Gallium) ~\le~ \left\{
  \begin{array}{l}
    72 \pm 12 \rm~SNU,~~~(no~oscillations), \\
    73 \pm 12 \rm~SNU,~~~(with~oscillations).
  \end{array}\quad\right.
\eeq{eq18}
Scaling up the \Beryllium\ neutrino bounds in Eq.~(\ref{eq14}) by the ratio
of the capture cross sections on \Gallium\ and \Chlorine, we find that the
bounds on the \Beryllium\ neutrino contribution to the \Gallium\ signals are:
\beq
  R(\Beryllium,\Gallium) ~<~ \left\{
  \begin{array}{rl}
     6.0 \rm~SNU, &\rm (no~oscillations), \\
    17.4 \rm~SNU, &\rm (with~oscillations),
  \end{array}\quad\right.
\eeq{eq19}
at the 95\% confidence level.  It will be interesting to test these bounds
by direct observation of the \Beryllium, or pp neutrinos themselves.

This work was supported in part by the U. S. Department of Energy grant
DE-FG05-92ER40691. The authors are grateful to Serguey Petcov and Gene Beier
for helpful conversations.  One of the authors (SPR) would like to thank
Prof.\ W. C. Haxton for the hospitality of the Institute for Nuclear Theory
at the University of Washington, where part of this work was carried out.

Note added: After this work was completed, the authors learned from Prof.\
David Schramm that he had obtained a bound in the non-oscillation case
similar to that in Eq.~(3).

\vskip2ex

\def\etal{\hbox{\it et al.}}
\def\ibid{\hbox{\it ibid.}}

\centerline{\bf References}
\begin{enumerate}

\item 
K. Hirata \etal, Phys.\ Rev.\ D {\bf 44}, 2241 (1991) and
Y. Suzuki, in {\it Frontiers of Neutrino Astrophysics}, Proceedings of the
   International Symposium on Neutrino Astrophysics, edited by Y. Suzuki and
   N. Nakamura (Universal Academic Press, Tokyo, 1993) p.~61.
For the most recent results see Y. Suzuki, 6th International Workshop on
   ``Neutrino Telescopes,'' February 22--24, 1994, Venice, Italy; and
International Topical Workshop on the Solar Neutrino Problem: Astrophysics
   or Oscillations? 28 February--1 March 1994, Laboratori Nazionali del Gran
   Sasso, L'Aquila, Italy.

\item 
J. N. Bahcall and M. Pinsonneault, Rev.\ Mod.\ Phys.\ {\bf 64}, 885 (1992);
J. N. Bahcall, {\it Neutrino Astrophysics} (Cambridge University Press,
   New York, 1993).

\item 
Ray Davis, in {\it Frontiers of Neutrino Astrophysics}, Proceedings of the
   International Symposium on Neutrino Astrophysics, edited by Y. Suzuki and
   N. Nakamura (Universal Academic Press, Tokyo, 1993) p.~47; and
6th International Workshop on ``Neutrino Telescopes,'' February 22--24, 1994,
   Venice, Italy; and
International Topical Workshop on the Solar Neutrino Problem: Astrophysics or
   Oscillations? 28 February--1 March 1994, Laboratori Nazionali del Gran
   Sasso, L'Aquila, Italy.

\item 
GALLEX Collaboration, P. Anselmann \etal, Report No.\ GX 44-1994 (submitted
to Physics Letters B, February 1994).

\item 
J. N. Abdurashitov \etal, Los Alamos Report No.\ LA-UR 94-1113 (submitted to
   Physics Letters B, April 1994).

\item 
J. N. Bahcall, {\it Neutrino Astrophysics} (Cambridge University Press,
   New York, 1993).

\item 
Reference 6, Section 8.2, p.\ 214-243.

\item 
For recent discussions of MSW solutions, see
S. Bludman, N. Hata, D. Kennedy, and P. Langacker, Phys.\ Rev.\ D {\bf 47},
   2220 (1993);
N. Hata and P Langacker, University of Pennsylvania Preprint No.\
   UPR-0592T (Nov.\ 1993);
P. I. Krastev and S. Petcov, Phys.\ Lett.\ B {\bf 299}, 99 (1993); and
L. Krauss, E. Gates, and M. White, Phys.\ Rev.\ Lett.\ {\bf 70}, 375 (1993).

\item 
See for example H. A. Bethe and J. N. Bahcall, Phys.\ Rev.\ D {\bf 44}, 2962
   (1991).

\item 
Reference 6, p.~209.

\item 
GALLEX Collaboration, P. Anselmann \etal, Phys.\ Lett.\ B {\bf 285}, 376
   (1992) and \ibid\ B {\bf 314}, 445 (1993).

\end{enumerate}

\vskip 2ex
\centerline{\bf Figure caption}

{\bf Fig.~1}.~~Normalized shapes of $\phi\sigma$ for various experiments.

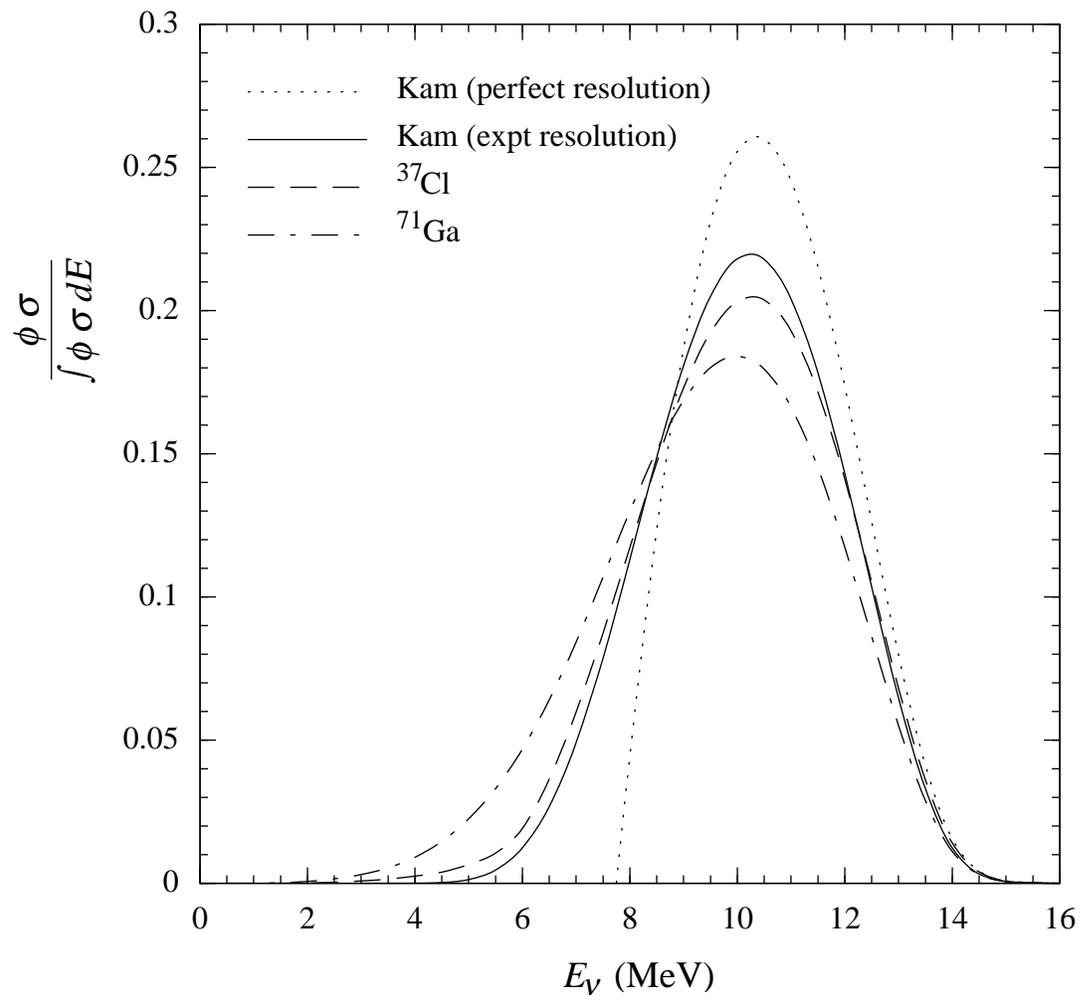
\end{document}